\documentclass[useAMS,usenatbib]{emulateapj}
\usepackage{comment}
\usepackage{xspace}
\usepackage{graphicx}
\usepackage{epsfig}
\usepackage[backref,colorlinks,citecolor=blue]{hyperref}

\providecommand{\url}[1]{\href{#1}{#1}}
\providecommand{\dodoi}[1]{doi:~\href{http://doi.org/#1}{\nolinkurl{#1}}}
\providecommand{\doeprint}[1]{\href{http://ascl.net/#1}{\nolinkurl{http://ascl.net/#1}}}
\providecommand{\doarXiv}[1]{\href{https://arxiv.org/abs/#1}{\nolinkurl{https://arxiv.org/abs/#1}}}

\newcommand{\Chandra}{\textit{Chandra}}
\newcommand{\nicer}{\textit{NICER}\xspace}

\shorttitle{Absorbed Flare in GRS 1915+105 with \nicer}
\shortauthors{Neilsen et al.}

\begin{document}

\title{A \textit{NICER} View of a  Highly-Absorbed Flare in GRS 1915+105}

\author{J.\ Neilsen\altaffilmark{1}}
\author{J.\ Homan\altaffilmark{2,3}}
\author{J.\ F.\ Steiner\altaffilmark{4}}
\author{G. Marcel\altaffilmark{1}}
\author{E.\ Cackett\altaffilmark{5}}
\author{R.\ A.\ Remillard\altaffilmark{4}}
\author{K.\ Gendreau\altaffilmark{6}}

\altaffiltext{1}{Villanova University, Department of Physics, Villanova, PA 19085, USA; jneilsen@villanova.edu}
\altaffiltext{2}{Eureka Scientific, Inc., 2452 Delmer Street, Oakland, CA 94602, USA}
\altaffiltext{3}{SRON, Netherlands Institute for Space Research, Sorbonnelaan 2, 3584 CA Utrecht, The Netherlands}
\altaffiltext{4}{MIT Kavli Institute for Astrophysics and Space Research, Cambridge, MA 02139, USA}
\altaffiltext{5}{Wayne State University, Department of Physics \& Astronomy, Detroit, MI 48201, USA}
\altaffiltext{6}{NASA Goddard Space Flight Center, Greenbelt, MD 20771, USA}

\begin{abstract}
After 26 years in outburst, the black hole X-ray binary GRS 1915+105 dimmed considerably in early 2018; its flux dropped sharply in mid-2019, and it has remained faint ever since. This faint period, the ``obscured state," is punctuated by occasional X-ray flares, many of which have been observed by \nicer as part of our regular monitoring program. Here we present detailed time-resolved spectroscopy of one bright flare, whose spectrum shows evidence of high column density partial covering absorption and extremely deep absorption lines (equivalent widths over 100 eV in some cases). We study the time-dependent ionization of the obscuring gas with {\sc xstar}, ultimately attributing the absorption to a radially-stratified absorber of density $\sim10^{12-13}$ cm$^{-3}$ at a $\sim$few$\times10^{11}$ cm from the black hole.  We argue that a vertically-extended outer disk could explain this obscuration. We discuss several scenarios to explain the obscured state, including massive outflows, an increase in the mass accretion rate, and changes in the outer disk that herald the approach of quiescence, but none are entirely satisfactory. Alternative explanations, such as obscuration by the accretion stream impact point, may be testable with current or future data.
\end{abstract}


\section{Introduction}
\label{sec:intro}

In 1992, the stellar mass black hole GRS 1915+105 was discovered as an X-ray transient with the WATCH instrument on the International Astrophysical Observatory ``GRANAT" \citep{CastroTirado92}. Black hole transients typically brighten quickly and fade slowly over the course of several months \citep{FBG04,Homan05b,McClintockRemillard06,Tetarenko16}, providing opportunities to study the physical processes involved in black hole accretion (e.g., relativistic jets, \citealt{Fender09}; ionized winds, \citealt{N13a}; and black hole spin, \citealt{McClintockShafee06}). But GRS 1915+105 did not fade in the months or years that followed, instead exhibiting an outburst so long and so bright that it has been debated whether it should be considered a persistent source instead of a transient \citep{Corral-Santana16,Tetarenko16,Deegan09}. Extensive studies with radio (e.g., \citealt{F99,K02}), infrared \citep{E98a,Rothstein05,Mikles06}, and X-ray telescopes \citep{MRG97,b00,l02,m08,nl09,n18} have revealed GRS 1915+105 to be an ideal target for studying the physics of jets, ionized winds, and variability around black holes, as well as the links between them (the `disk-wind-jet connection`). 

\begin{figure}
\includegraphics[width=3.2in,clip=true,trim=2 5 0 0]{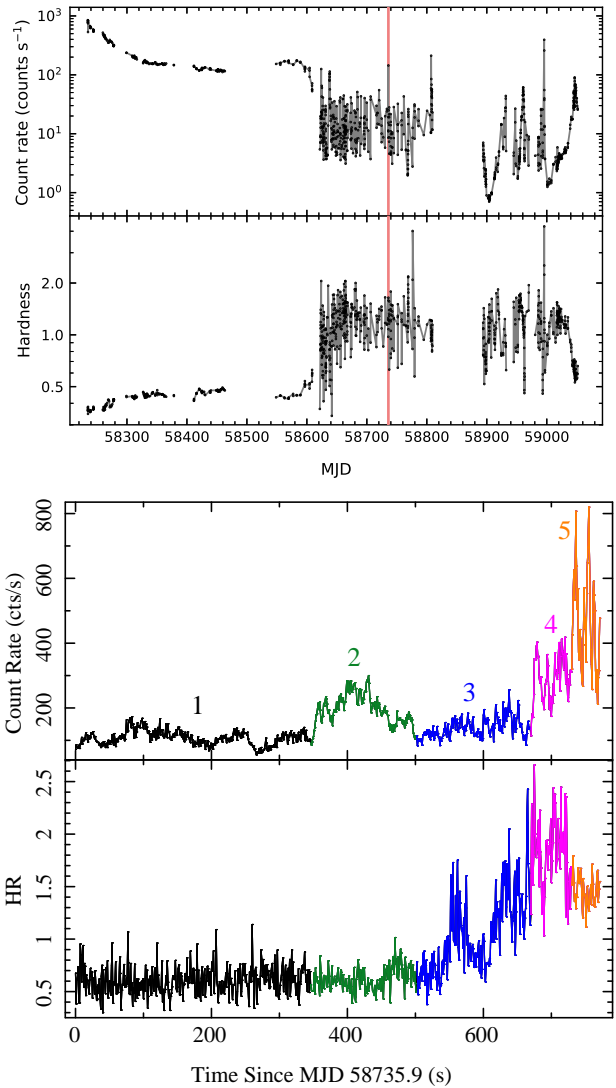}\\
\includegraphics[width=3.2in]{f1b.eps}
\caption{Top panels: long-term \nicer lightcurve and hardness (4--10 keV/0.5--4 keV) of GRS 1915+105, showing the approach to the obscured state and its subsequent variability. The location of the 2019 September 9 flare is indicated with a vertical red line. Bottom panels: 1-s lightcurve and hardness ratio HR for the flare. HR is defined as the ratio of the 5--12 keV and 2--5 keV count rates. Strong spectral evolution is apparent.\label{fig:lc}}
\end{figure}

After 26 years of strong erratic variability, early 2018 brought a sudden change for GRS 1915+105: its X-ray brightness decreased exponentially \citep{Negoro18} (see Figure \ref{fig:lc}). In 2019 April and May, around MJD 58600, the flux dropped again over several weeks \citep{Homan19}, reaching tens of \nicer counts s$^{-1}$ (compared to $\lesssim4000$ counts s$^{-1}$ in 2017 and early 2018; \citealt{n18}). The flux decrease coincided with a significant hardening of the X-ray spectrum. Contemporaneous \Chandra, \textit{NuSTAR}, and \textit{Swift} spectra showed strong emission and absorption lines and evidence of Compton-thick obscuration \citep{Miller19a,Miller19b}. Together, these facts indicate that in 2019 GRS 1915+105 entered a new state from which it has not yet emerged: the ``obscured" state (\citealt{Miller20}).

Transient or variable obscuration is not unusual. \citet{Miller19a} suggested a parallel to X-ray ``changing-look" AGN, where obscuration of the central engine can vary from Compton-thick to Compton-thin, and which have been known for $\sim$two decades (\citealt{Guainazzi02,Matt03} and references therein; see also \citealt{Risaliti02}). These variations occur on timescales of months to years, corresponding to tens to hundreds of seconds around stellar mass black holes. Indeed, column density variations on such short timescales have been seen in several black hole X-ray binaries (e.g., V404 Cyg, \citealt{Motta17a,Motta17b}, see also \citealt{Oosterbroek97,Zycki99};  V4641 Sgr, \citealt{Revnivtsev02,Morningstar14}; and Swift J1858.6-014, \citealt{Hare20}). Several of these sources were studied in detail with \textit{RXTE} and \textit{NuSTAR} by \citet{Koljonen20}. The origin of these obscurers may vary from X-ray binaries to AGN, but it is clear that they provide important diagnostics of the environments of accreting black holes.

In X-ray binaries with compact objects, epochs with strong, variable local absorption are often associated with high-amplitude flaring activity  (e.g., \citealt{Walton17,Hare20,Munoz-Darias20}). The obscured state in GRS 1915+105 is no exception: its lightcurve indicates numerous flares with significant absorption variability and strong line features. In this paper, we discuss the spectral evolution of one such flare observed with \nicer (Figure \ref{fig:lc}). We present our data analysis in Section \ref{sec:obs}, simple spectral models in Section \ref{sec:spec}, and models of the X-ray absorber in \ref{sec:xstar}. We discuss our conclusions in Section \ref{sec:discuss}.

\setcounter{footnote}{0}

\section{Observations and Data Reduction}
\label{sec:obs}

\nicer observed GRS 1915+105 on 2019 September 9 from UTC 15:18:40 to 23:23:20, with a total good time exposure of approximately 5200 s (ObsID 2596012703). Using {\sc heasoft} v6.26 with the 20190516 {\sc caldb}, we applied a custom set of screening criteria  to all NICER data, omitting detectors 14, 34, and 54 due to their propensity for increased noise (e.g., \citealt{Bogdanov19}). Good time intervals were selected from periods outside the SAA in which (a) pointing was aligned within 1.5 arcmin of GRS 1915+105, (b) all subsystems were configured normally for science, and (c) there were no data dropouts. We further required that the spacecraft is oriented with SUN\_ANGLE $>60 \degr$  (or when in shadow,  SUNSHINE=0), with ELV $> 20\degr$ and BR\_EARTH $> 30\degr$.

The resulting active detectors were screened on their X-ray, overshoot, and undershoot rates; any detector outside the median by $>15$ MAD (median absolute deviation) was rejected. We used the fast-to-slow chain PI-ratio as suggested in the \nicer data analysis guide\footnote{https://heasarc.gsfc.nasa.gov/docs/nicer/data\_analysis/nicer\_\\analysis\_guide.html} to discriminate background events via the standard ``trumpet'' energy filter. We used {\sc xselect} to extract 1-second lightcurves from the filtered data in two energy bands (2--5 keV and 5--12 keV) and used the ratio of the count rates in the hard band and soft band to define a hardness ratio HR. 

\section{Time-Resolved Spectroscopy}
\label{sec:spec}

The first $\sim$20 ks of this observation show secular variability and a low count rate, near 20 cts s$^{-1}$. But shortly after UTC 21:36, one GTI showed a strong ($\sim8\times$) flare that exhibited significant spectral hardening (see Figure \ref{fig:lc}). For the remainder of this paper, we restrict our attention to this GTI and the spectral evolution of the flare. The flare lightcurve in Figure \ref{fig:lc} shows several intervals with distinct properties. Given the strong variability, we skip analysis of the average spectrum and report our time-resolved analysis.

\begin{figure*}
\centerline{\includegraphics[width=\textwidth]{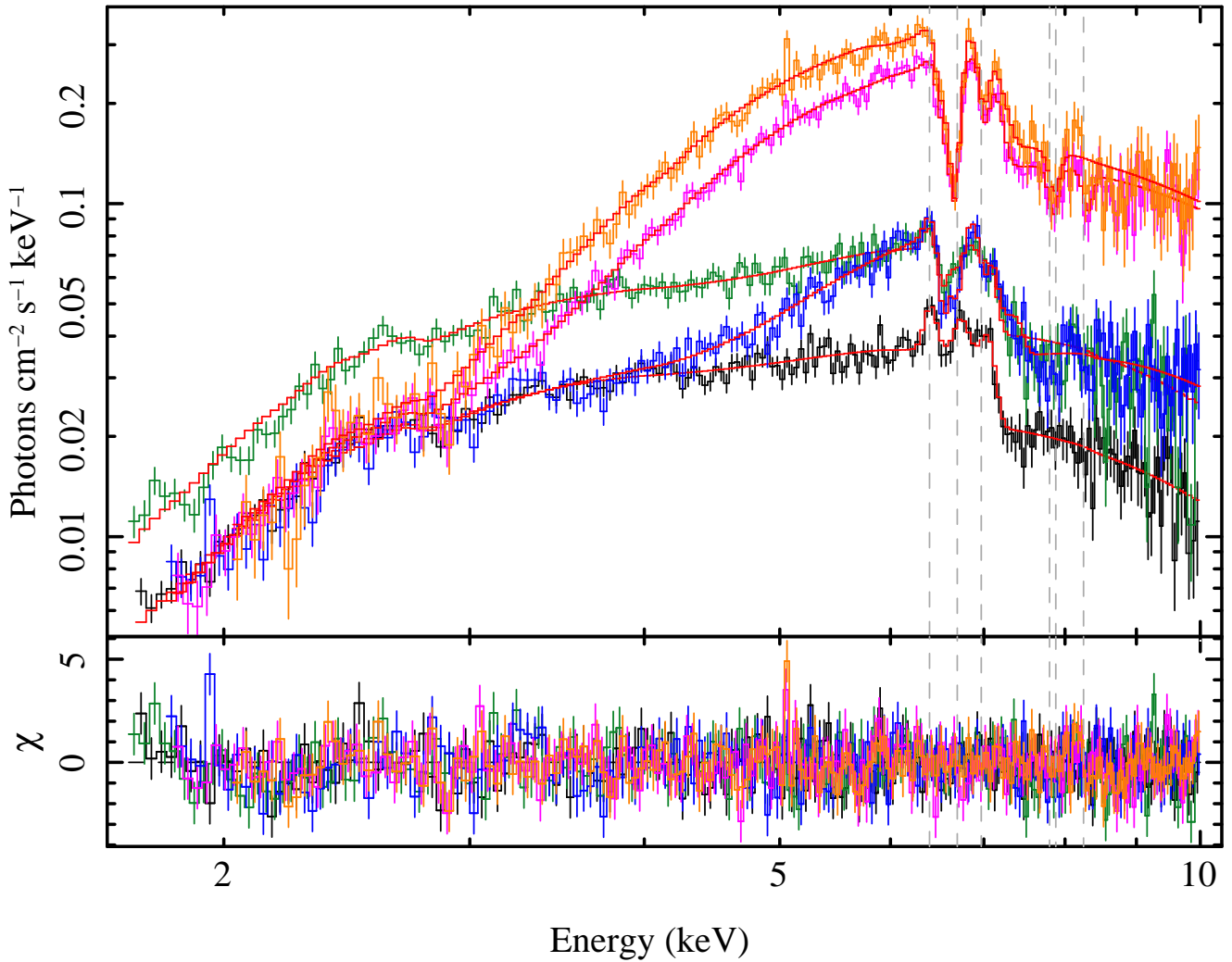}}
\caption{Binned \nicer spectra of the five flare intervals; colors match Figure \ref{fig:lc}. The spectra show significant hardening as the flux rises, along with the appearance of deep absorption lines in the 6.5-7 keV region. Residuals (bottom panel) are shown for the {\tt nthcomp} continuum model with individual Gaussian components to model the absorption lines. Vertical dashed gray lines show the rest energies of some identified lines and line candidates: 6.4 keV, 6.7 keV, 6.97 keV, 7.8 keV, 7.88 keV, and 8.25 keV. \label{fig:spec}}
\end{figure*}

\subsection{Lightcurve and Spectra} 
We divide the observation into five separate intervals based solely on changes in the count rate. Interval 1 consists of the first $333$ seconds of the observation, when the count rate is slowly varying below $\sim200$ counts s$^{-1}$ and HR is constant but noisy. The count rate of interval 1 is well above that of the previous GTI. The count rate rises and falls by a factor of 2-3 over the $146$ s of interval 2, but there are no corresponding changes in HR. During interval 3 (158 s), HR rises rapidly while the count rate shows only modest increases. The count rate rises sharply in interval 4 (99 s), which cover the remainder of the observation; its spectrum is effectively the average spectrum of the flare. Interval 5 is the final 43 s of the observation, when the count rate is highest. Note that intervals 4 and 5 are not completely independent.

These intervals cleanly partition the spectral evolution of this flare. We used {\sc xselect} to extract spectra from each interval. Background spectra for each interval were computed using tracers of in-band contamination from particle events: the cutoff rigidity COR\_SAX value, along with per-detector rates for overshoot and undershoot events. A minute-interval Gaussian smoothing of these quantities was compared against look-up tables of the same based on \nicer observations of Rossi X-ray Timing Explorer background fields (e.g., Remillard et al., in preparation) and used to compute a model background lightcurve and its corresponding spectrum. Typical differences between various background models are statistically negligible when the source-to-background ratio is $>10,$ as is the case here.

\defcitealias{Miller20}{M20}

The resulting background-subtracted spectra are shown in Figure \ref{fig:spec}. The spectrum of interval 1, shown in black, is unusually flat, with a strong edge near 7 keV and several weak emission lines. There is no significant change in the continuum shape during interval 2 (green), although the flux is higher than interval 1. This is as expected from the lightcurve and the time-dependent HR shown in Figure \ref{fig:lc}. Several smaller changes are apparent in Figure \ref{fig:spec}, however: the 7 keV edge is slightly shallower than in interval 1, and a strong absorption line appears between 6.5 and 7 keV. The width of this feature suggests a blend of multiple lines, most likely highly ionized Fe (see also \citealt{Miller20}, hereafter \citetalias{Miller20}). The weak emission lines appear to be present as well.

During interval 3 (blue), the soft X-ray flux is suppressed to the level of interval 1, while the hard X-ray emission is comparable to that of interval 2, and the 7 keV edge is even more shallow. This evolution is likely a result of two competing factors: a reduction of the count rate coupled with a hardening of the spectrum. This is characteristic of an increase in both the luminosity and the absorbing column density. At the same time, the ionized absorption lines increase in depth.

The flare itself, intervals 4 (magenta) and 5 (orange), is much brighter but the spectra have similarities to the previous intervals. As in interval 3, the continuum is effectively identical to that of interval 1 below 3 keV. Above 3 keV, the spectrum rises smoothly up to the Fe K region, where a number of strong absorption lines and a relatively shallow edge at 7 keV are apparent. Intervals 4 and 5 are very similar, though the spectrum of interval 5 is brighter between 3 and 6 keV and slightly steeper above 8 keV.

 We perform all our spectral analysis in the Interactive Spectral Interpretation System ({\sc isis}) v.1.6.2-43 (\citealt{HD00}) with {\sc heasoft} v.6.26; we assume a distance of 8.6 kpc \citep{Reid14}. To reduce oversampling of the energy resolution, the 1.5--10 keV spectra are rebinned by a factors of 3, 4, and 5 from 1.5--3 keV, 3--6 keV, and 6--10 keV, respectively.
 
\subsection{Spectral Model}

These model-independent findings hint at the presence of local absorption in GRS 1915+105 that varies over time not only in column density but also in ionization state. As we demonstrate below and in Section \ref{sec:xstar}, the data are well described with a series of variable partially-covered absorbers. In this section, we consider models for the cold absorption and the underlying continuum using individual Gaussians for the narrow emission and absorption lines.

We explore a number of typical models for the X-ray continuum, including power laws with and without cutoffs, disk blackbodies, and Comptonization, and various combinations of the same with varying absorption, but generally find that these models either do not describe the data or produce unphysical parameters. The spectra are quite hard and there is no sign of the typical bright, hot disk component in GRS 1915+105 \citep{Peris16}.

The similarity of the continuum shape to X-ray spectra of V404 Cyg, V4641 Sgr, and Swift J1858.6-014, as well as typical Seyfert 2 X-ray spectra (e.g., \citealt{Moran01}) motivates our choice of a partially covered absorber (we use {\tt tbnew}, with abundances and cross-sections from \citet{Wilms00} and \cite{Verner96}, respectively). Two continuum models prove most effective: {\tt nthcomp} \citep{Zdziarski96,Zycki99} and {\tt bbody}. For {\tt nthcomp}, we use a blackbody seed photon spectrum whose temperature is free to vary, and we require the electron temperature $kT_{\rm e}>$ 1 keV. 

In addition to the narrow lines, we find good improvement to our fits by including an edge at $\sim2.75$ keV. This energy does not correspond to the K-edge rest energy of any abundant element (or any H-like or He-like ion), so this feature may represent a residual calibration uncertainty in our data.

Our final model for each observation therefore consists of: either {\tt nthcomp} or {\tt bbody} for the underlying continuum, a {\tt tbnew} component tied across all observations to represent the ISM column density, a second partially-covered {\tt tbnew} component to represent the local obscuring medium, an edge, a broad iron line ({\tt rellline}; \citealt{Dauser10}), three Gaussian emission lines, and several additional Gaussian absorption lines. In light of the apparent presence of the 6.5--7 keV emission lines during intervals 2 and 3, we fit a single set of narrow emission lines jointly to all five observations.

Two of the strongest absorption lines appear at 6.7 and 7 keV and are clearly identifiable as Fe\,{\sc xxv} He$\alpha$ and Fe\,{\sc xxvi} Ly$\alpha.$ But there is also significant absorption at $6.4-6.5$ keV, and it is difficult to fit the full absorption line profile, even accounting for the included emission lines, without a second line. We therefore include an additional line at 6.5 keV, which we tentatively identify as a blend of absorption features from less-ionized species of Fe, but it could also be an indicator of the velocity structure of the ionized obscuration (see \citetalias{Miller20}). Observations of similar states at higher spectral resolution may be able to disentangle these effects. Other lines are detectable in some intervals at $\sim7.85$ keV and $\sim8.3$ keV. Both the Fe\,{\sc xxv} He$\beta$ line  and the Ni\,{\sc xxvii} He$\alpha$ line are located near 7.8 keV, and the Fe\,{\sc xxvi} Ly$\beta$ line is located at 8.26 keV. We suggest these as likely IDs for these features.

We fit these models to the data using standard fit minimization procedures. The {\tt nthcomp} model has 84 free parameters and produces a best-fit reduced $\chi^2$ of 1.19 ($\chi^2/\nu=1025/859$), while the {\tt bbody} model (74 free parameters) has reduced $\chi^2$ of 1.21 ($\chi^2/\nu=1053/869$). Given the large number of parameters and possible correlations between them, we estimate our uncertainties using an {\sc isis} implementation\footnote{We use the {\tt emcee} routine from the ECAP/Remeis Observatory/MIT ISISscripts available here: http://www.sternwarte.uni-erlangen.de/isis/.} of the Markov Chain Monte Carlo (MCMC) routine {\sc emcee-hammer} \citep{ForemanMackey13}. For both models we use flat priors and 10 walkers per free parameter. The {\tt bbody} model has a total of 740 walkers; we allow each walker to evolve for 10,000 steps, reaching 7.4 million samples. Defining the burn-in period as the time it takes the minimum, maximum, and median fit statistic to reach steady state, we discard the initial 4,500 steps for each parameter , drawing results from the remaining 5,500 steps (4.07 million samples). With 84 parameters, the {\tt nthcomp} converges somewhat more slowly, and we use a total of 840 walkers $\times$ 11,400 steps. Again we take results from the final 5,500 steps for each walker, totaling 4.62 million samples. Following the {\tt emcee} documentation\footnote{https://emcee.readthedocs.io/en/stable/tutorials/autocorr/}, we estimate the typical autocorrelation times $\tau$ for these runs to be roughly 430 steps, which means our effective sample size $N/\tau$ is $\gtrsim9500$ for both models.

To compute errors on fit parameters, we find the minimum-width 90\% credible interval, which involves calculating the cumulative distribution of all the samples for a given parameter and identifying the narrowest interval that contains 90\% of the samples. When this interval includes the minimum (maximum) sample value, we set the corresponding endpoint of the credible interval equal to the lower (upper) limit for the parameter. For derived quantities (e.g., equivalent width, mass, radius, etc.), we draw 1000 samples from the posterior distribution, calculate the derived quantity for each of the 1000 samples, and use the same algorithm as above to determine the credible interval.

\subsection{Results}

As noted above, the {\tt nthcomp} and {\tt bbody} models produce very similar $\chi^2_\nu$; {\tt bbody} is a slightly worse fit but has fewer free parameters, and the difference is only $\Delta\chi^2\approx3$ per free parameter. For studying the continuum evolution, we treat these models as equally viable. In the discussion that follows, we refer to the {\tt nthcomp} model for interval 1 as N1, and use B1 as shorthand for the {\tt bbody} model for interval 1, and so on. The full continuum parameters are quoted in Table \ref{spectable}.

The ISM column density determined by these fits, $N_{\rm H,ISM}=4.6^{+0.2}_{-0.1}\times10^{22}$ cm$^{-2}$ for {\tt nthcomp} and $N_{\rm H,ISM}=5.4^{+0.2}_{-0.1}\times10^{22}$ cm$^{-2}$ for {\tt bbody}, is comparable to previously-measured values for GRS 1915+105 (e.g., \citealt{l02}). The two models return statistically indistinguishable results for the edge as well: $(E_{\rm edge},\tau_{\rm edge})=(2.75\pm0.04$ keV, $0.15\pm0.04)$ for {\tt nthcomp} and (2.78$^{+0.01}_{-0.07}$ keV, 0.18$^{+0.01}_{-0.07}$) for {\tt bbody}.

\begin{figure}[h]
\centerline{\includegraphics[width=0.49\textwidth]{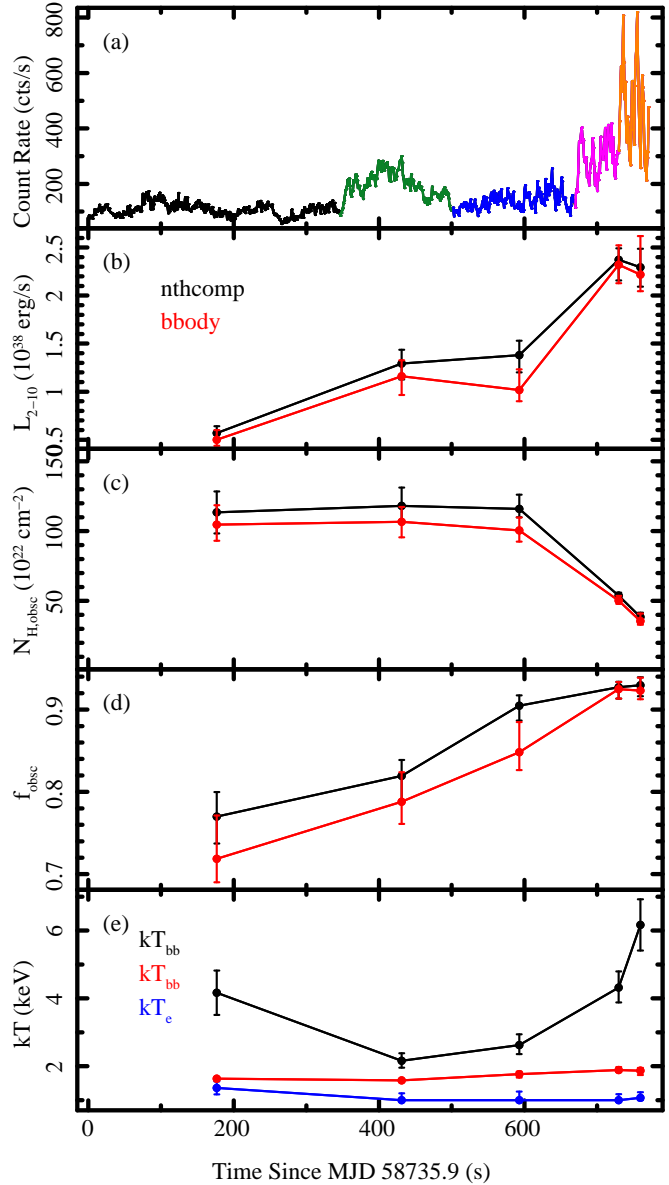}}
\caption{Continuum parameters for the {\tt nthcomp} (black) and {\tt bbody} (red) models. Panels show the following: (a) \nicer lightcurve, for reference; (b) 2--10 keV luminosity; (c) cold obscuring column density; (d) partial covering fraction $f_{\rm obsc}$; (e) blackbody temperature (with {\tt nthcomp} electron temperature in blue). See text for details. \label{fig:contpars}}
\end{figure}

The remainder of the parameters are time-dependent, and we illustrate their evolution in Figure \ref{fig:contpars} with the flare lightcurve in panel (a) for reference. In panel (b), we show the inferred intrinsic/unabsorbed 2--10 keV luminosity of the continuum component. Both models suggest a mostly-monotonic rise through interval 4 with a slight but not statistically significant decrease in interval 5\footnote{We reiterate that interval 5 is the latter half of interval 4.}.  Because the count rate of interval 5 is higher than the average rate of interval 4, it is evident that a significant part of the evolution of the flare must be driven by changes in the local obscuring medium.

This fact is also evident in the parameters of this obscuration. Panels (c) and (d) show the column density $N_{\rm H,obsc}$ and partial covering fraction $f_{\rm obsc}$ over time. Both models indicate a large cold column density $N_{\rm H,obsc}\approx110\times10^{22}$ cm$^{-2}$ through interval 3 that drops to $\sim$52$\times10^{22}$ cm$^{-2}$ and $\sim$37$\times10^{22}$ cm$^{-2}$ during intervals 4 and 5, respectively. Meanwhile, $f_{\rm obsc}$ rises smoothly from $\sim0.75$ to $\sim0.93$. These results indicate that the flare is the combined result of a $\sim$4$\times$ increase in the intrinsic luminosity of GRS 1915+105 and a $\sim$3$\times$ decrease in the column density of cold obscuring gas. As we shall see in Section \ref{sec:xstar}, this decrease cannot be entirely attributed to the ionization of the obscuring medium.
 
The {\tt nthcomp} model requires an unusually hot seed photon spectrum, with $kT_{\rm bb}=4.2\pm0.7$ keV for N1, which drops down to $2.2\pm0.2$ keV in N2 and then rises to $6.2\pm0.8$ keV in N5 (see panel e). Meanwhile, the electron temperature $kT_{\rm e}$ is very low ($1-1.6$ keV). The photon index $\Gamma$ ranges from 1.2 to 2.4, which implies a high optical depth to Compton scattering (see the discussion in \citealt{N16} and references therein). When we instead model the emission as a pure 
(unscattered) blackbody, the inferred temperature varies slowly from $\sim1.6$ keV to 1.9 keV. Given the luminosity and temperature, the implied blackbody radius (assuming a spherical emitter) is $\sim10^6$ cm, which is comparable to the gravitational radius of the black hole $r_g\approx1.8\times10^6$ cm. We take this as evidence of a compact X-ray source in either case.

For the broad iron line, we fix the inclination to $i=66^\circ$ and the spin to $a=0.998$ \citep{McClintockShafee06}, and we assume an emissivity index $q=3$. The line energies range from $5.72$ keV (N4, B4) to 6.4 keV (B3), and the inferred radii range from $1.6R_{\rm g}$ (N3) to $23R_{\rm g}$ (B3). The lines as modeled contribute $\sim10-15\%$ of the 2--10 keV luminosity, but given their low energies it is not clear if they are physical or represent un-modeled effects of Compton scattering (see Section \ref{sec:assumptions} and \citealt{Motta17a} for more details). 

Finally, the absorption lines are visibly strong, but a quantitative comparison to another source is illuminating. During its 2005 outburst, GRO J1655--40 exhibited some of the strongest Fe absorption lines in a stellar mass black hole published to date (\citealt{M06a,m08}; these had equivalent widths of $\gtrsim30$ eV, according to \citealt{N12b}). In contrast, the equivalent width of the Fe\,{\sc xxv} absorption line peaks at $123$ eV in N5, where the total equivalent width of the detected Fe absorption features is $\sim300$ eV.  In general, the lines are unresolved; many are blueshifted but this is often not statistically significant on a line-by-line basis. Rather than report the properties and time dependence of all these features in detail, we give as an example the line parameters for N4 and B4 in Table \ref{linetable}. Given that we apply more physical photoionization models to study the absorber in Section \ref{sec:xstar}, this sample of parameters should be sufficient to give a sense of the lines.

\begin{deluxetable*}{lcccccccccc}
\tabletypesize{\scriptsize}
\tablewidth{0pt}
\tablecaption{Continuum Spectroscopy of the Absorbed Flare
\label{spectable}}
\tablehead{\colhead{Parameter} & N1&N2&N3&N4&N5&B1&B2&B3&B4&B5}
\startdata
$N_{\rm H,ism}$\tablenotemark{c}&$4.6_{-0.1}^{+0.2}$ & 4.6\tablenotemark{a} & 4.6\tablenotemark{a} & 4.6\tablenotemark{a} & 4.6\tablenotemark{a} & $5.4_{-0.1}^{+0.2}$ & 5.4\tablenotemark{a} & 5.4\tablenotemark{a} & 5.4\tablenotemark{a} & 5.4\tablenotemark{a}  \vspace{1mm}\\
$N_{\rm H,obsc}$\tablenotemark{c} &$110_{-20}^{+10}$ & $120\pm10$ & $116_{-6}^{+10}$ & $54_{-3}^{+2}$ & $38\pm3$ & $100\pm10$ & $110\pm10$ & $101_{-8}^{+9}$ & $50_{-2}^{+3}$ & $36_{-3}^{+5}$  \vspace{1mm}\\
$f_{\rm obsc}$ &$0.77^{+0.03}_{-0.03}$ & $0.82_{-0.03}^{+0.02}$ & $0.9_{-0.02}^{+0.01}$ & $0.93^{+0.01}_{-0.01}$ & $0.93^{+0.01}_{-0.01}$ & $0.72_{-0.03}^{+0.05}$ & $0.79_{-0.03}^{+0.04}$ & $0.85_{-0.02}^{+0.04}$ & $0.92^{+0.01}_{-0.01}$ & $0.92_{-0.01}^{+0.02}$  \vspace{1mm}\\
$K$\tablenotemark{c} &$0.07^{+0.01}_{-0.01}$ & $0.18^{+0.03}_{-0.03}$ & $0.16_{-0.03}^{+0.04}$ & $0.17^{+0.03}_{-0.03}$ & $0.19^{+0.04}_{-0.04}$ & $0.08^{+0.01}_{-0.01}$ & $0.19_{-0.02}^{+0.03}$ & $0.18_{-0.02}^{+0.03}$ & $0.42_{-0.02}^{+0.03}$ & $0.4_{-0.02}^{+0.05}$  \vspace{1mm}\\
$kT_{\rm bb}$ (keV) &$4.2^{+0.7}_{-0.7}$ & $2.2^{+0.2}_{-0.2}$ & $2.6^{+0.3}_{-0.3}$ & $4.3_{-0.4}^{+0.5}$ & $6.2^{+0.8}_{-0.8}$ & $1.63_{-0.06}^{+0.04}$ & $1.58_{-0.06}^{+0.04}$ & $1.8^{+0.1}_{-0.1}$ & $1.89_{-0.06}^{+0.09}$ & $1.9^{+0.1}_{-0.1}$  \vspace{1mm}\\
$kT_{\rm e}$ (keV) &$1.4\pm0.2$ & $1^{+0.2}$\tablenotemark{b} & $1^{+0.3}$\tablenotemark{b} & $1^{+0.2}$\tablenotemark{b} & $1.1_{-0.1}^{+0.2}$ & \nodata & \nodata & \nodata & \nodata & \nodata \vspace{1mm}\\
$\Gamma$ &$1.2\pm0.2$ & $2.4_{-0.5}^{+0.3}$ & $2.1_{-0.4}^{+0.2}$ & $2_{-0.3}^{+0.1}$ & $1.6_{-0.2}^{+0.1}$ & \nodata & \nodata & \nodata & \nodata & \nodata  \vspace{1mm}\\
$-\log(f_{\rm Fe})\tablenotemark{c}$ &\nodata & \nodata &$8.9^{+0.01}_{-0.01}$ & $8.62_{-0.08}^{+0.07}$ & $8.45_{-0.06}^{+0.08}$ & \nodata & \nodata &$9^{+0.01}_{-0.01}$ & $8.62_{-0.09}^{+0.05}$ & $8.45_{-0.06}^{+0.09}$  \vspace{1mm}\\
$E_{\rm Fe}$ (keV) &\nodata & \nodata &$5.9_{-0.13}^{+0.06}$ & $5.72_{-0.04}^{+0.12}$ & $6.09_{-0.11}^{+0.09}$ & \nodata & \nodata &$6.4_{-0.09}^{+0.13}$ & $5.72_{-0.04}^{+0.11}$ & $6\pm0.1$  \vspace{1mm}\\
$R_{\rm in} (R_{\rm ISCO})$ &\nodata & \nodata &$1.3_{-0.3}^{+2}$ & $3.4_{-0.9}^{+2.4}$ & $11_{-3}^{+2}$ & \nodata & \nodata &$19_{-4}^{+5}$ & $3.4_{-0.9}^{+2.2}$ & $9\pm3$  \vspace{1mm}\\
$E_{\rm edge}$ (keV) &$2.75_{-0.04}^{+0.04}$ & 2.75\tablenotemark{a} & 2.75\tablenotemark{a} & 2.75\tablenotemark{a} & 2.75\tablenotemark{a} & $2.78_{-0.07}^{+0.01}$ & 2.78\tablenotemark{a} & 2.78\tablenotemark{a} & 2.78\tablenotemark{a} & 2.78\tablenotemark{a}  \vspace{1mm}\\
$\tau_{\rm edge}$ &$0.15_{-0.04}^{+0.04}$ & 0.15\tablenotemark{a} & 0.15\tablenotemark{a} & 0.15\tablenotemark{a} & 0.15\tablenotemark{a} & $0.18_{-0.07}^{+0.01}$ & 0.18\tablenotemark{a} & 0.18\tablenotemark{a} & 0.18\tablenotemark{a} & 0.18\tablenotemark{a}  \vspace{1mm}\\
$L_{\rm 2-10,38}$ &$0.57_{-0.07}^{+0.07}$ & $1.3_{-0.2}^{+0.1}$ & $1.4_{-0.2}^{+0.1}$ & $2.4_{-0.2}^{+0.1}$ & $2.3_{-0.2}^{+0.2}$ & $0.5_{-0.06}^{+0.1}$ & $1.2_{-0.2}^{+0.2}$ & $1_{-0.1}^{+0.2}$ & $2.3_{-0.2}^{+0.2}$ & $2.2_{-0.2}^{+0.4}$  \vspace{1mm}
\enddata
\tablecomments{Continuum and broad iron line parameters for the five intervals of the \nicer flare. Columns N1-N5 report parameters for the {\tt nthcomp} continuum model for intervals 1-5, respectively; columns B1-B5 show the same parameters for the {\tt bbody} model. $N_{\rm H,ism}$ and $N_{\rm H,obsc}$ are the ISM and obscuring column densities; $f_{\rm obsc}$ is the partial covering fraction of the obscuring medium. $K$ is the normalization of the {\tt nthcomp} or {\tt bbody} model. $kT_{\rm bb}$ and $kT_{\rm e}$ represent the blackbody and electron temperatures, respectively. $f_{\rm Fe}$ is the iron line flux; $E_{\rm Fe}$ is its energy, and $R_{\rm in}$ is the inner radius of the line region. $E_{\rm edge}$ and $\tau_{\rm edge}$ give the location and depth of the additional edge. $L_{2-10,38}$ is the unabsorbed 2--10 keV luminosity in units of $10^{38}$ erg s$^{-1}.$ Errors represent minimum-width 90\% credible intervals. See text for details.}
\tablenotetext{a}{This parameter is tied across all five intervals.}
\tablenotetext{b}{This parameter is pegged at its lower limit.}
\tablenotetext{c}{Column densities are quoted in units of 10$^{22}$ cm$^{-2}$. $f_{\rm Fe}$ is measured in units of erg s$^{-1}$ cm$^{-2}$.}
\end{deluxetable*}

\begin{deluxetable*}{lccccccc}
\tabletypesize{\scriptsize}
\tablewidth{0pt}
\tablecaption{Line Parameters During Interval N4\label{linetable}}
\tablehead{\colhead{Line ID} & Type & $E_0$ & Norm & $E_{\rm obs}$ & $v$ & EW & $\sigma$\\&&(keV)& ($10^{-3}$ ph/s/cm$^2$)&(keV)& (km/s) & (eV) & (eV)}
\startdata
Fe\,K$\alpha$ & E & 6.40 & $2_{-0.6}^{+0.5}$ & $6.41_{-0.02}^{+0.01}$ & $-400_{-700}^{+900}$ & $8_{-2}^{+2}$ & $<0.7$  \vspace{1mm}\\
Fe\,{\sc xxv} He$\alpha$ & E & 6.70 & $1.8_{-0.7}^{+0.8}$ & $6.74_{-0.02}^{+0.02}$ & $-2000_{-1000}^{+1000}$ & $6_{-2}^{+3}$ & $40_{-10}^{+20}$  \vspace{1mm}\\
Fe\,{\sc xxvi} Ly$\alpha$ & E & 6.97 & $1.3_{-0.5}^{+0.9}$ & $7.06_{-0.03}^{+0.03}$ & $-4000_{-1000}^{+1000}$ & $5_{-2}^{+3}$ & $40$\tablenotemark{a}  \vspace{1mm}\\
Fe\,{\sc $<$xxv}\tablenotemark{c} & A & \nodata\tablenotemark{c} & $14_{-1}^{+5}$ & $6.54_{-0.02}^{+0.02}$ & \nodata\tablenotemark{c} & $51_{-5}^{+15}$ & $<7$  \vspace{1mm}\\
Fe\,{\sc xxv} He$\alpha$ & A & 6.70 & $26_{-2}^{+3}$ & $6.6704_{-0.0006}^{+0.0231}$ & $1330_{-1040}^{+30}$ & $97_{-12}^{+6}$ & $<7$\tablenotemark{b}  \vspace{1mm}\\
Fe\,{\sc xxvi} Ly$\alpha$ & A & 6.97 & $18_{-4}^{+3}$ & $7.014_{-0.017}^{+0.009}$ & $-1900_{-400}^{+700}$ & $63_{-14}^{+7}$ & $<7$\tablenotemark{b}  \vspace{1mm}\\
Fe\,{\sc xxv}/Ni\,{\sc xxvii}\tablenotemark{c} & A & \nodata\tablenotemark{c} & $6_{-2}^{+2}$ & $7.85_{-0.03}^{+0.03}$ & \nodata\tablenotemark{c} & $50_{-20}^{+10}$ & $<7$\tablenotemark{b}  \vspace{1mm}\\
Fe\,{\sc xxvi} Ly$\beta$ & A & 8.26 & $5_{-3}^{+3}$ & $8.3_{-0.03}^{+0.04}$ & $-2000_{-2000}^{+1000}$ & $40_{-20}^{+20}$ & $<7$\tablenotemark{b}
\enddata
\tablecomments{Example line parameters for interval N4 to illustrate the features detected during the flare and their relative strengths. Emission and absorption lines are categorized by E and A, respectively. $E_0$ and $E_{\rm obs}$ are the line rest energy and observed energy in keV. Norm is the integrated line flux in $10^{-3}$ photons s$^{-1}$ cm$^{-2}$. $v$ is the Doppler shift in km s$^{-1}.$ EW with the line equivalent width in eV; $\sigma$ is the physical line width in eV. Errors represent minimum-width 90\% credible intervals.}
\tablenotetext{a}{The Fe\,{\sc xxv} (6.7 keV) and Fe\,{\sc xxvi} (6.97 keV) emission line widths are tied.}
\tablenotetext{b}{The absorption line widths are tied.}
\tablenotetext{c}{The 6.54 keV feature is likely a blend of lines from ionization states below Fe\,{\sc xxv}. The 7.85 keV feature is likely a blend of the Fe\,{\sc xxv} He$\beta$ line (7.88 keV) and the Ni\,{\sc xxvii} He$\alpha$ line (7.80 keV). Without specific line IDs, we do not quote rest energies or Doppler shifts for these features.}
\end{deluxetable*}

\section{Photoionization Modeling}
\label{sec:xstar}

With viable models for the underlying variability of GRS 1915+105 during this flare, we turn our attention to characterizing the ionized absorber. For this analysis, we use {\sc xstar} \citep{Kallman01,Bautista01} to compute electron energy level populations for various ionizing spectra and use these to fit the \nicer spectra directly with the analytic model {\tt warmabs} (see \citealt{Kallman09}).

In order to compute the level populations and  photoionization models, we must first specify the ionizing radiation field. This is straightforward when the emission geometry is simple and the absorber is clearly optically thin. In the case of this flare from GRS 1915+105, however, the geometry of emission and absorption is complicated. The intrinsic emission is relatively hard, possibly Compton thick, and likely compact, but it is modified by a highly inhomogeneous absorber whose location and properties are unknown. Yet the relative positions of the ionized and cold obscuration have significant implications for the ionizing radiation field.

Here we consider the question in the context of two limiting cases: is the cold gas interior or exterior to the hot gas? If the ionized absorber lies inside the cold absorber---between it and the black hole---then it is appropriate to use the intrinsic X-ray spectrum of the source, i.e., the pure unabsorbed {\tt nthcomp} or {\tt bbody} model, to calculate the ionization balance of the hot gas. But if the hot gas is farther from the black hole than the cooler obscuring medium, then it is heated not by the intrinsic X-ray spectrum of GRS 1915+105, but instead by the partially-absorbed radiation field that emerges from behind the cold gas. This is discussed more in Section \ref{sec:geom}, but the upshot of these considerations is that for each interval with strong absorption lines (2-5), we have four possible radiation fields: {\tt nthcomp} or {\tt bbody}, with or without the cold partial covering absorber detailed in Table \ref{spectable}. 
For each of these four ionizing spectra in intervals 2-5, we compute {\sc xstar} level populations over the ionization parameter (\citealt{Tarter69}) range $\log\xi=2-4.$ We run {\sc xstar} assuming a constant electron density of $10^{14}$ cm$^{-3}$ and an initial temperature of $4\times10^6$ K, though we set {\tt niter}=99 so that the code computes thermal equilibrium in each radial zone. We explore the density further in Section \ref{sec:geom}, so we consider this value to be a placeholder (the lines studied here are not density-sensitive). In order to ensure the calculations are done in the optically-thin limit, we use a low column density ($10^{17}$ cm$^{-2}$). \citetalias{Miller20} studied a different wind in GRS 1915+105 during the onset of the obscured state (MJD 58603.19) using {\sc spex} models that do not require an optically thin absorber and found similar results to ours, so we do not believe the {\sc xstar} assumptions lead to significant bias.  

With the 16 level population files in hand we proceed to fit the \nicer spectra using {\tt warmabs}. We remove the Gaussian absorption lines from the models described in Section \ref{sec:spec} and replace them with a two-zone ionized absorber with the same partial covering fraction as the cold absorber. In effect, we treat the full absorption (hot and cold gas) as a single, radially-stratified medium; whether the temperature increases outwards or inwards is a direct consequence of using an absorbed or unabsorbed continuum, respectively, to compute the {\sc xstar} files.

To preserve the {\sc xstar}-calculated ionization balance while we fit the {\tt warmabs} models to the data, we freeze the underlying continuum parameters, with two exceptions: the cold column density $N_{\rm H,obsc}$ and the partial covering fraction $f_{\rm obsc}$. As there may be small trade-offs between the cold and hot absorption, we allow these to vary. We note again that in our ionization analysis, this is the covering fraction for both the hot and cold gas; see below for more details. For each dataset and {\tt warmabs} run, we are left with 9 free parameters: $N_{\rm H,obsc}$ and $f_{\rm obsc}$, a turbulent velocity for the ionized gas $v_{\rm turb}$, plus column densities $N_{\rm H,warm/hot}$, ionization parameters ($\xi_{\rm warm/hot}$), and Doppler shifts ($z_{\rm warm/hot}$=$v_{\rm warm/hot}$/c) for each zone. 

{\tt warmabs} can be slow to fit and compute confidence intervals directly, but we note that our {\tt emcee} burn-in period is typically a short $\sim100$ steps for these models. Since MCMC is  designed to sample the posterior distributions of the parameters and not to maximize the goodness of fit, we develop an iterative fitting procedure to find the best-fit parameters. This involves: (1) initializing the parameters to reasonable values, (2) fitting the model to the data, (3) performing a 100-step {\tt emcee} run\footnote{We use the same setup as in Section \ref{sec:spec}: 10 walkers per free parameter and flat priors.}, (4) using the best parameters from the resulting chain as initial parameters for a new fit if $\chi^2$ improves by $>2$, (5) repeating steps (3) and (4) until $\Delta\chi^2<2$, and (6) performing a final 1000-step {\tt emcee} run, from which we draw our results. The autocorrelation time estimated using the method described above is roughly 50 steps, so we have $\sim1800$ independent samples (90 walkers $\times1000$ steps / 50).

Our first quality check for each ionizing radiation field (absorbed vs unabsorbed, {\tt nthcomp} vs {\tt bbody}) is the total $\chi^2$ for intervals 2-4. As before, we use shorthand to refer to the models and intervals: the ionizing spectrum for AN2 is the absorbed, partially-covered {\tt nthcomp} spectrum for interval 2, while UN2 uses the unabsorbed {\tt nthcomp} spectrum for the same interval, and so on. For reference, the models described in Section \ref{sec:spec}, which fit the absorption lines with independent Gaussians, return $\chi^2_{\tt nthcomp}=823$ and $\chi^2_{\tt bbody}=853.$ Relative to these, the best-performing {\tt warmabs} model is AN (the absorbed {\tt nthcomp} continuum), with $\chi^2_{\rm AN}=841;$ the unabsorbed {\tt nthcomp} models UN are a close second at $\chi^2_{\rm UN}=844.$ The blackbody models are not as successful ($\chi^2_{\rm AB}=869,$ $\chi^2_{\rm UB}=874$). We note that we also fit models where the partial covering only applied to the cold absorption, i.e., a covering fraction of 1 for the {\tt warmabs} components. The values of $\chi^2$ for these models were between 40 and 125 higher than their partially-covered counterparts, strongly supporting  partial covering for the ionized absorption. Given the consistently smaller $\chi^2$ values returned by the {\tt nthcomp} models, we focus on these in the discussion that follows. 

\begin{figure}
\centerline{\includegraphics[width=0.49\textwidth]{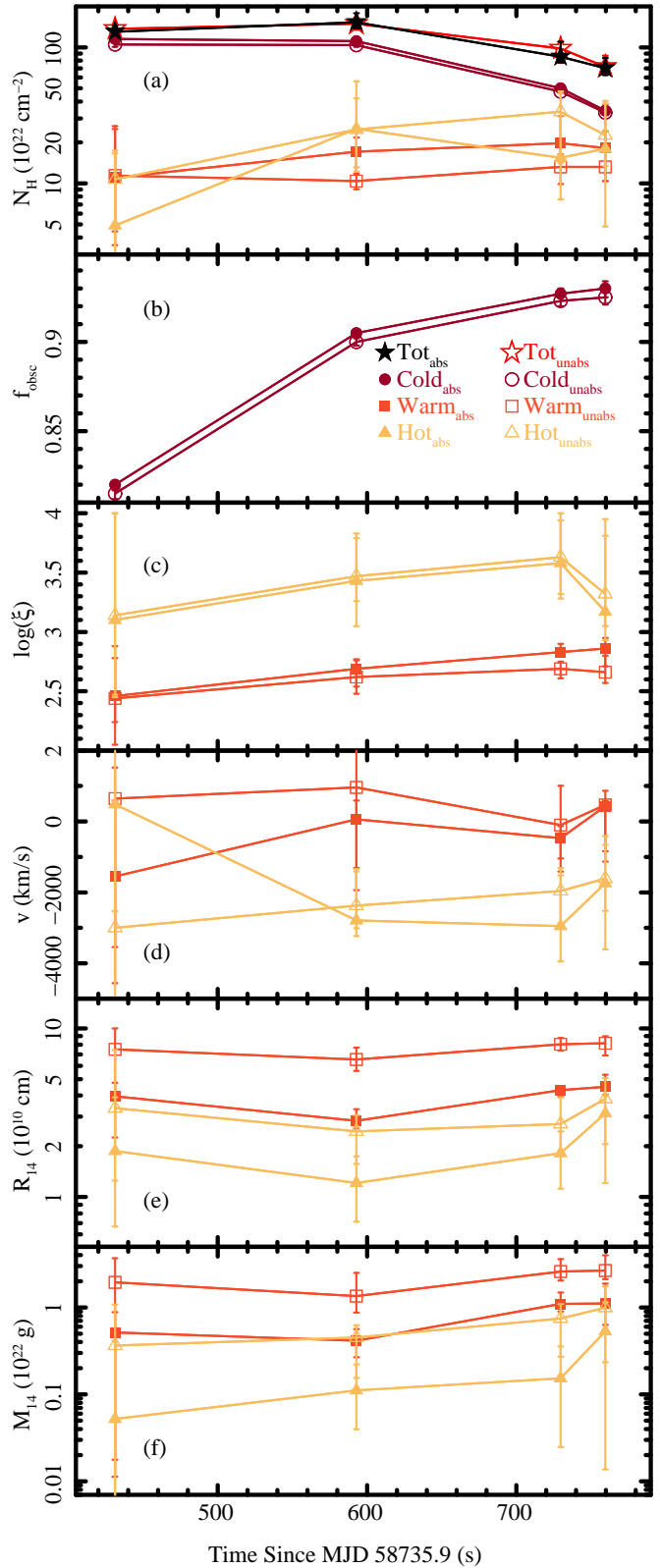}}
\caption{Parameters of the partially ionized partial covering obscuring medium during intervals 2-4. Filled symbols represent AN ionization models and open symbols represent the UN ionization models. Panels show: (a) Total (stars), cold (circles), and ionized (squares and triangles) column densities; (b) partial covering fraction; (c) ionization parameters; (d) Doppler shifts; (e) radii, assuming a compact X-ray source and a density of $10^{14}$ cm$^{-3}$, and (f) shell mass, also assuming a density of $10^{14}$ cm$^{-3}$. In panels (a) and (b), we have offset the UN values downward slightly because they are indistinguishable from the AN values.\label{fig:windpars}}

\end{figure}

Our results for the AN and UN models are tabulated in Table \ref{windtable}, and the evolution of several key parameters is shown in Figure \ref{fig:windpars}. The evolution of the cold obscuring medium is not distinguishable from the phenomenological fits with Gaussians: $N_{\rm H,obsc}$ is steady around $\sim$110$\times10^{22}$ cm$^{-2}$ in intervals 2 and 3, then falls to $\sim$35$\times10^{22}$ in interval 5. At the same time, $f_{\rm obsc}$ rises from 0.82 to 0.93. These values are consistent with our continuum models in Table \ref{spectable}. This is not particularly surprising as the other continuum parameters are frozen, but it suggests that the ionized absorption and the cold absorption are separable: if the hot gas contributed significantly to the continuum shape (i.e., beyond the absorption lines), then replacing the Gaussians with {\tt warmabs} would also require modifications to the underlying spectral model.

Broadly, the {\tt warmabs} components are characterized by relatively low ionization parameters ($\log(\xi)=2.4-3.6$; $\log(\xi)\gtrsim4$ is more typical for disk winds; \citealt{N13a}; \citetalias{Miller20}). This is less an indication of weakly-ionized gas and more an indication of the X-ray spectral hardness: for this spectrum these are the ionization parameters where the ion fractions of Fe\,{\sc xxv} and Fe\,{\sc xxvi} are significant. The column densities are large, usually $\gtrsim10^{23}$ cm$^{-2}.$ Owing to \nicer's energy resolution, the Doppler shifts are not tightly constrained. However, the two components are generally distinguishable in ionization, and the less ionized component $\log(\xi)=2.5-3$ usually has a smaller Doppler shift. The hotter ($\xi_{\rm hot}$) component appears to be outflowing in some intervals. There are systematic gain uncertainties at 7 keV that can be up to 10-20 eV (effectively $\sigma_v\lesssim850$ km s$^{-1}$; this represents a significant  improvement in the calibration since \citealt{n18}). But in many cases, the velocity of the hotter component and the velocity difference between the two ionized components are larger than this systematic uncertainty in the gain. Thus, to the extent that their Doppler shifts can be distinguished statistically, we can conclude that these two components are distinct. The hot component is likely outflowing, while the warm component is consistent with being at rest. 

\begin{deluxetable*}{lcccccccc}
\tabletypesize{\scriptsize}
\tablewidth{0pt}
\tablecaption{{\sc xstar} Models of the Ionized Absorbers
\label{windtable}}
\tablehead{\colhead{Parameter} & AN2&AN3&AN4&AN5&N2&N3&N4&N5}
\startdata
$N_{\rm H,obsc}$\tablenotemark{a} &$115_{-5}^{+5}$ & $111_{-3}^{+3}$ & $50.1_{-1.7}^{+0.4}$ & $34.2_{-1.6}^{+0.8}$ & $114_{-4}^{+6}$ & $113_{-3}^{+2}$ & $51.4_{-0.9}^{+0.8}$ & $36_{-1.2}^{+0.8}$  \vspace{1mm}\\
$f_{\rm H,obsc}$\tablenotemark{a} &$0.82_{-0.003}^{+0.002}$ & $0.905_{-0.002}^{+0.002}$ & $0.927_{-0.002}^{+0.003}$ & $0.93_{-0.005}^{+0.004}$ & $0.82_{-0.003}^{+0.002}$ & $0.905_{-0.002}^{+0.002}$ & $0.928_{-0.002}^{+0.002}$ & $0.93_{-0.004}^{+0.004}$  \vspace{1mm}\\
$N_{\rm H,warm}$\tablenotemark{a} &$11_{-7}^{+15}$ & $17_{-5}^{+5}$ & $20_{-3}^{+12}$ & $18_{-8}^{+20}$ & $11_{-7}^{+14}$ & $10_{-1}^{+7}$ & $13_{-3}^{+6}$ & $13_{-3}^{+11}$  \vspace{1mm}\\
$\log(\xi_{\rm warm})$ &$2.5_{-0.2}^{+0.4}$ & $2.69_{-0.15}^{+0.08}$ & $2.83_{-0.04}^{+0.07}$ & $2.86_{-0.14}^{+0.09}$ & $2.4_{-0.4}^{+0.3}$ & $2.6_{-0.1}^{+0.1}$ & $2.69_{-0.08}^{+0.06}$ & $2.66_{-0.09}^{+0.14}$  \vspace{1mm}\\
$v_{\rm warm}$ (km/s) &$-2000_{-2000}^{+2000}$ & $100_{-2000}^{+500}$ & $-500_{-900}^{+300}$ & $400_{-1600}^{+100}$ & $600_{-5200}^{+900}$ & $1000_{-2000}^{+2000}$ & $-100_{-900}^{+1100}$ & $500_{-1300}^{+400}$  \vspace{1mm}\\
$N_{\rm H,hot}$\tablenotemark{a} &$5_{-4}^{+11}$ & $30_{-10}^{+30}$ & $15_{-8}^{+32}$ & $20_{-10}^{+20}$ & $11_{-10}^{+7}$ & $20_{-10}^{+20}$ & $30_{-20}^{+10}$ & $23_{-9}^{+16}$  \vspace{1mm}\\
$\log(\xi_{\rm hot})$ &$3.1_{-0.6}^{+0.9}$ & $3.4_{-0.4}^{+0.4}$ & $3.6_{-0.3}^{+0.4}$ & $3.2_{-0.2}^{+0.8}$ & $3.1_{-0.7}^{+0.9}$ & $3.5_{-0.2}^{+0.4}$ & $3.6_{-0.4}^{+0.3}$ & $3.3_{-0.3}^{+0.5}$  \vspace{1mm}\\
$v_{\rm hot}$ (km/s) &$500_{-3000}^{+100}$ & $-2800_{-400}^{+1500}$ & $-3000_{-1000}^{+1400}$ & $-2000_{-2000}^{+1000}$ & $-3000_{-7000}^{+10000}$ & $-2400_{-600}^{+1000}$ & $-2000_{-900}^{+700}$ & $-1600_{-900}^{+900}$  \vspace{1mm}\\
$v_{\rm turb}$ (km/s) &$<7000$ & $1300_{-700}^{+1200}$ & $1100_{-700}^{+400}$ & $900_{-400}^{+1100}$ & $<6800$ & $1000_{-400}^{+1500}$ & $900_{-400}^{+800}$ & $1400_{-600}^{+1000}$  \vspace{1mm}\\
$N_{\rm H,tot}$\tablenotemark{a} &$130_{-6}^{+13}$ & $150_{-20}^{+20}$ & $85_{-8}^{+24}$ & $70_{-8}^{+13}$ & $140_{-10}^{+10}$ & $150_{-10}^{+20}$ & $100_{-20}^{+10}$ & $72_{-8}^{+15}$  \vspace{1mm}\\
$L_{38}$ &$0.45$ & $0.39$ & $1.26$ & $1.46$ & $1.56$ & $1.78$ & $3.14$ & $3.06$ \vspace{1mm}\\
$R_{14,\rm warm}$\tablenotemark{b} &$4_{-1.7}^{+0.8}$ & $2.8_{-0.3}^{+0.5}$ & $4.3_{-0.3}^{+0.2}$ & $4.5_{-0.5}^{+0.8}$ & $8_{-3}^{+2}$ & $6.5_{-1}^{+1.1}$ & $8.1_{-0.7}^{+0.7}$ & $8.2_{-1.3}^{+0.8}$ \vspace{1mm}\\
$\Delta R_{14,\rm warm}$\tablenotemark{b} &$0.11_{-0.1}^{+0.07}$ & $0.17_{-0.06}^{+0.04}$ & $0.2_{-0.05}^{+0.1}$ & $0.2_{-0.1}^{+0.2}$ & $0.11_{-0.08}^{+0.09}$ & $0.1_{-0.02}^{+0.06}$ & $0.13_{-0.04}^{+0.05}$ & $0.13_{-0.04}^{+0.1}$ \vspace{1mm}\\
$M_{14,\rm warm}$\tablenotemark{b} &$0.5_{-0.5}^{+0.4}$ & $0.4_{-0.1}^{+0.1}$ & $1.1_{-0.2}^{+0.4}$ & $1.1_{-0.5}^{+0.8}$ & $2_{-2}^{+2}$ & $1.4_{-0.5}^{+1.2}$ & $2.6_{-0.5}^{+1}$ & $2.7_{-0.5}^{+1.3}$ \vspace{1mm}\\
$M_{14,\rm warm}/\Delta t$\tablenotemark{b} &$70_{-60}^{+50}$ & $17_{-6}^{+6}$ & $29_{-5}^{+11}$ & $30_{-10}^{+20}$ & $300_{-200}^{+200}$ & $60_{-20}^{+50}$ & $70_{-10}^{+30}$ & $70_{-10}^{+30}$ \vspace{1mm}\\
$R_{14,\rm hot}$\tablenotemark{b} &$2_{-1}^{+2}$ & $1.2_{-0.5}^{+0.5}$ & $1.8_{-0.7}^{+0.6}$ & $3.1_{-1.9}^{+0.8}$ & $3_{-2}^{+4}$ & $2.5_{-0.9}^{+0.6}$ & $2.7_{-0.9}^{+1.2}$ & $4_{-2}^{+1}$ \vspace{1mm}\\
$\Delta R_{14,\rm hot}$\tablenotemark{b} &$0.05_{-0.04}^{+0.1}$ & $0.3_{-0.2}^{+0.2}$ & $0.2_{-0.1}^{+0.2}$ & $0.2_{-0.2}^{+0.1}$ & $0.11_{-0.1}^{+0.07}$ & $0.2_{-0.1}^{+0.2}$ & $0.34_{-0.21}^{+0.1}$ & $0.2_{-0.1}^{+0.1}$ \vspace{1mm}\\
$M_{14,\rm hot}$\tablenotemark{b} &$0.05_{-0.05}^{+0.36}$ & $0.11_{-0.07}^{+0.11}$ & $0.2_{-0.1}^{+0.2}$ & $0.5_{-0.5}^{+0.6}$ & $0.4_{-0.4}^{+0.7}$ & $0.5_{-0.3}^{+0.2}$ & $0.7_{-0.5}^{+0.3}$ & $1_{-0.8}^{+0.8}$ \vspace{1mm}\\
$M_{14,\rm hot}/\Delta t$\tablenotemark{b} &$7_{-7}^{+46}$ & $5_{-3}^{+5}$ & $4_{-3}^{+5}$ & $10_{-10}^{+20}$ & $50_{-50}^{+90}$ & $19_{-12}^{+7}$ & $20_{-13}^{+9}$ & $20_{-20}^{+20}$ \vspace{1mm}\\
$\chi^2/$d.o.f &224.18/181 & 233.25/181 & 202.78/180 & 180.92/174 & 223.39/181 & 233.59/181 & 204.74/180 & 182.70/174  \vspace{1mm}
\enddata
\tablecomments{{\sc xstar} photoionization models of absorption in intervals 2-5. Columns AN2-AN5 report parameters for an absorbed ionizing continuum, while  columns N2-B5 show the same parameters assuming the absorption is ionized by the unabsorbed continuum. $N_{\rm H,obsc}$ and $f_{\rm obsc}$ are as in Table \ref{spectable}. $N_{\rm H,warm}$ and $N_{\rm H,hot}$ are the column densities of ionized absorption components 1 and 2, respectively; $\xi^1$ and $\xi^2$ are their ionization parameters and $v_{\rm warm}$ and $v_{\rm hot}$ are their Doppler shifts. $v_{\rm turb}$ is the turbulent velocity of the absorber (tied across both components). $N_{\rm H,tot}$ is the total column density. $R_{14},$ $\Delta R_{14}$, and $M_{14}$ are the radius, radial extent, and mass of the absorber assuming a uniform density of $10^{14}$ cm$^{-3}$. $\Delta t_n$ is the time between the end of interval 1 and the midpoint of interval $n$. See text for details. Errors represent minimum-width 90\% credible intervals.}
\tablenotetext{a}{Column densities are quoted in units of 10$^{22}$ cm$^{-2}$.}
\tablenotetext{b}{Radii are given in units of $10^{10}$ cm. Masses are given in units of 10$^{22}$ g, and $M/\Delta t$ is quoted in units of $10^{18}$ g s$^{-1}.$
}
\end{deluxetable*}

One of the strengths of photoionization analysis is its ability to reveal the location of absorbing/emitting gas, assuming the accretion geometry is understood (see relevant caveats in \citealt{N16}). Here the continuum emission appears to be compact (recall from Section \ref{sec:spec} that the inferred blackbody radius is comparable to the radius of the event horizon). For this discussion, we assume it is reasonable to treat the emission as a point source. We use the ionization parameter defined by \citet{Tarter69}: \begin{equation}\xi=\frac{L}{n R^2},\label{eq:xi}\end{equation} where $L$ is the ionizing luminosity, $n$ is the electron density, and $R$ is the distance from the source. Given $L$ and $n$, we can calculate \begin{equation}R=\sqrt{\frac{L}{n\xi}}.\label{eq:r}\end{equation}In Table \ref{windtable} we quote radii for a density $n=10^{14}$ cm$^{-3}$ as $R_{14,\rm warm}$ for $\xi_{\rm warm}$ and $R_{14,\rm hot}$ for $\xi_{\rm hot}$. To calculate these radii, we use bolometric luminosities, which can be calculated directly from the model normalization for {\tt bbody}. The bolometric correction is $\sim20-30$\%. For {\tt nthcomp} we approximate the bolometric luminosity by assuming the same bolometric correction for the 2--10 keV flux as the {\tt bbody} model). For the AN models, we use the apparent luminosity of the partially-covered continuum; for the UN models we use the intrinsic luminosity of the unabsorbed continuum.

Regardless of the model, $R_{14}$ lies between $10^{10}$ cm and $10^{11}$ cm for both zones, which is comparable to the scale of some disk winds previously studied in GRS 1915+105 (e.g., \citealt{N11a,N12a}). Assuming the density is uniform, the radial extent of ionized absorber should be \begin{equation}\Delta R=N_{\rm H}/n.\label{eq:dr}\end{equation} Using the same notation as for $R$, we find $\Delta R_{14}$ between $\sim0.02R_{14}$ and $0.25R_{14}.$ If these values are accurate, they suggest the absorption lines originate in relatively narrow shells far from the black hole. A higher density would imply a smaller radius with a smaller $\Delta R/R,$ i.e., a relatively narrower shell.

Because of the ambiguity in the velocity structure of the wind, it is not possible to calculate a mass loss rate directly (see examples in \citealt{l02,U09}). However, the column density and radius provide an estimate of the total mass of the ionized (spherical) shell: \begin{equation}\label{eq:m}M=4\pi R^2N_{\rm H}m=4\pi m\frac{L N_{\rm H}}{n \xi},\end{equation} where $m=2.4\times10^{-24}$ g is the average atomic mass per hydrogen atom (see \citealt{N11a} and references therein). For a density of $10^{14}$ cm$^{-3}$, $M_{14}$ ranges from $5\times10^{20}$ g to $2.7\times10^{22}$ g in these zones; this is large but (a) it is still a tiny fraction of the likely mass of the disk \citep[][and references therein]{Deegan09} and (b) the uncertainties are large. Finally, $M$  divided by the time since interval 1, $\Delta t$, gives an upper limit on the average rate of change of the shell mass (for a fixed density). The values quoted here range from 0.2-3$\times$ the Eddington rate for a radiative efficiency of 10\%, though again is not clear how much mass is being lost here.  If the column density of the ionized absorption is non-zero during interval 1 (despite the emission line spectrum and the non-detection of absorption lines), or if the density were higher (both $M$ and $M/\Delta t$ are inversely proportional to density), this would imply a smaller value and rate of change for the shell mass. On the other hand, if the absorption arises in rotating structures in the outer disk (see Sections \ref{sec:geom} and \ref{sec:variability}), variations in $M$ may simply represent azimuthal asymmetries in the obscuring structures.

We can also explore changes in these quantities directly. Remarkably, there is only marginal evidence for changes in the ionized absorption. The hot gas column densities do not show significant variability; this is best seen in panel (a) of Figure \ref{fig:windpars},  where---to within errors---the total column density (hot plus cold gas, shown as stars) is roughly $N_{\rm H,obsc}$ plus a constant $\sim20-40\times10^{22}$ cm$^{-2}.$ To the extent that these are constant, it suggests that any changes in the individual line strengths should be attributed to ionization variations and the changing shape of the incident spectrum. Given the size of the error bars, there are no clear trends in the velocity or column density of the two ionized components.

The ionization parameters (panel c) do appear to rise during the flare, which is suggestive of a correlation with luminosity if it is real. For the cooler component,  the increase is comparable to what would be expected for a constant value of $nR^2$ ($\Delta\log\xi\sim0.25$ between intervals 2 and 4); the increase is slightly larger for the hotter component, which would imply a slight decrease in $nR^2$. There are also hints of a possible decline in $\xi$ during interval 5, but given the short $\sim40$ s exposure time, it is difficult to determine precisely.

The derived quantities ($R,$ $\Delta R$, $M$, and $M/\Delta t$) track the underlying observable parameters ($N_{\rm H},$ $L,$ and  $\xi$); where variations in the observables seem to be significant, the derived quantities appear to vary as well. For instance, between AN3 and AN4, the combined effect of the increase in luminosity, column density, and ionization in the cooler {\tt warmabs} component is a statistically-significant increase in its inferred mass. But as with the observables described above, there are no clear or strong trends that emerge throughout the flare. 

\section{Discussion}
\label{sec:discuss}

In 2019, GRS 1915+105 entered a deep low state---the ``obscured" state---after decades of intense activity. An extensive monitoring campaign with \nicer (PI: Neilsen) has revealed a number of bright flares that punctuate the low-level variability of the obscured state. In this paper, we have explored the spectral variability of one such flare that occurred in 2019 September (see Figure \ref{fig:lc}).

The spectra (Figure \ref{fig:spec}) are unusually hard for GRS 1915+105, with no evidence of the bright, hot disk emission that dominated much of its spectral variability in the last 25 years (e.g., \citealt{B97b,N11a}). Instead, we find a deep edge at 7 keV, Fe emission lines at low flux, and extremely strong absorption lines at high flux. Transitions between emission- and absorption-line spectra are not new for GRS 1915+105 (\citealt{nl09}), even on short timescales \citep{U10}. But the high equivalent widths of these features and the context in which they appear is quite unusual. Here we discuss the origin of these features and their implications. 

\subsection{Absorption Geometry}
\label{sec:geom}
Our analysis of the \nicer spectra of the flare using {\sc xstar} in conjunction with standard continuum modeling indicates that the absorption in GRS 1915+105 is well described by an inhomogeneous obscuring medium spanning a wide range of temperatures. The cold portion, modeled with {\tt tbnew}, has a column density that peaks near $1.2\times10^{24}$ cm$^{-2}.$ The hot ($>10^6$ K) portion is consistent with two components that differ in ionization and velocity, with a total column density near $3\times10^{23}$ cm$^{-2}$ that is relatively steady once the flare begins (i.e., interval 2 and beyond). 

There are, however, fundamental uncertainties in our photoionization analysis: we do not know the geometry of the absorbers or their densities (the absorption lines studied here are only sensitive to density in the very high density limit, $n\gtrsim10^{17}$ cm$^{-3}$). We have already established these issues separately above by (1) distinguishing models where the {\tt warmabs} components are ionized by the absorbed continuum (AN/AB) or the unabsorbed continuum (UN/UB) and (2) quoting the parameters of the absorbers with their densities explicitly noted. As we explain here, these uncertainties are directly related to the ionization and temperature profile of the gas.

In Table \ref{windtable}, we give the radius and width of each absorber for a density of $10^{14}$ cm$^{-3}$. These are generally well separated (see panel (e) of Figure \ref{fig:windpars}). For example, during interval AN4, $R_{14,\rm warm}=4.3_{-0.3}^{+0.2}\times10^{10}$ cm $\approx2.3\times10^4$ $r_g$ and $\Delta R_{14,\rm warm}=0.20_{-0.05}^{+0.09}\times10^{10}$ cm, while $R_{14,\rm hot}=1.8_{-0.7}^{+0.6}\times10^{10}$ cm and $\Delta R_{14,\rm hot}=0.2_{-0.1}^{+0.2}\times10^{10}$ cm. In effect, these are two completely distinct narrow shells, which does not seem like a particularly probable configuration for the absorbers if they have the same origin. In principle, this is what we might expect for a two-phase medium, with different temperature gases in pressure equilibrium (e.g., \citealt{Krolik81,Chakravorty13}). However, both our zones are on the hot branch of the thermodynamic stability curve, so it does not appear that the hot and warm gas are colocated.

\begin{figure*}
\centerline{\includegraphics[angle=90,clip=true,trim=60 20 30 20,width=\textwidth]{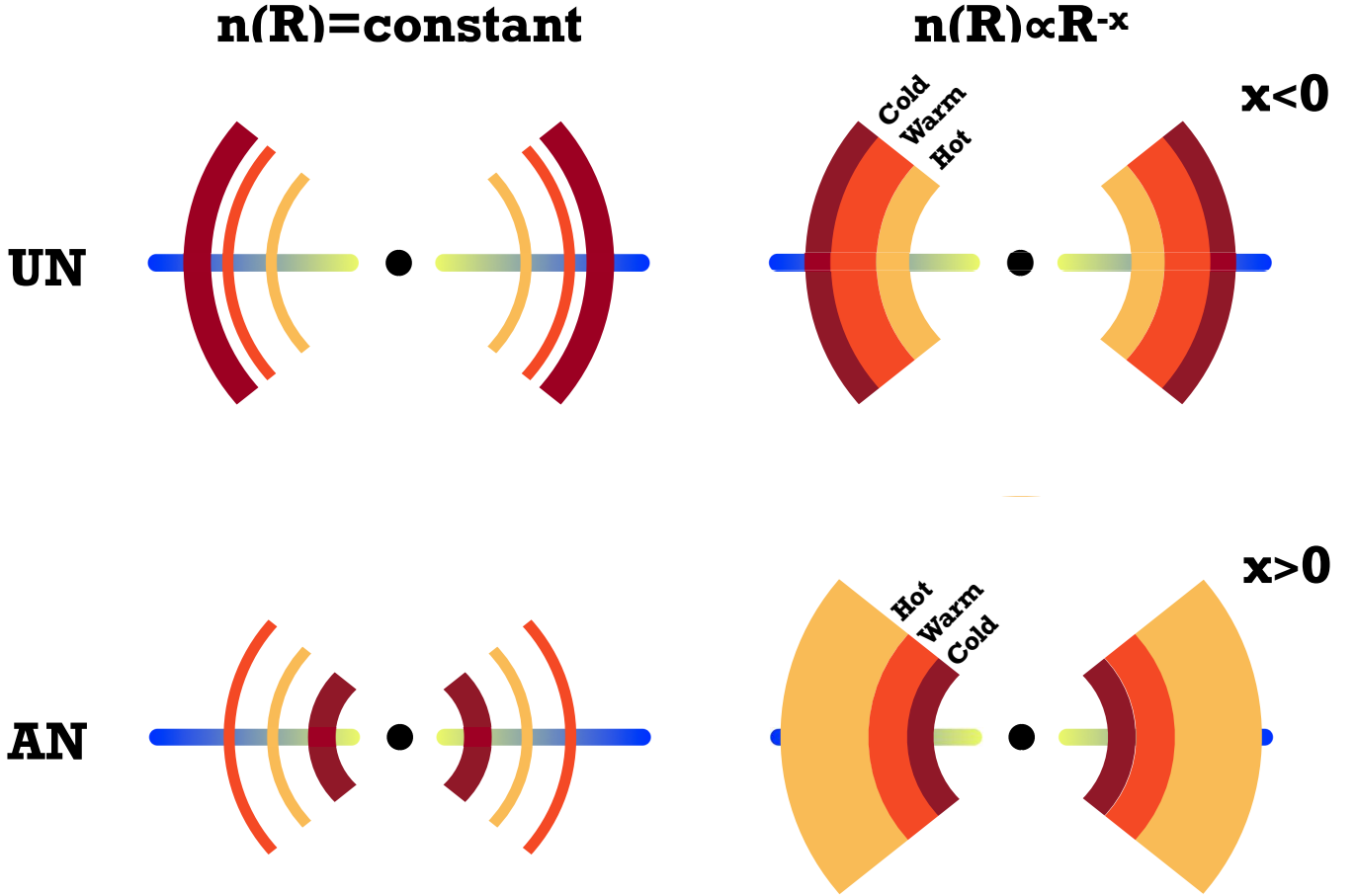}}
\caption{Cartoon of the absorption geometry for various scenarios described in Section \ref{sec:geom}. The cold, warm, and hot absorbers are shown in brown, red, and orange, respectively. The left column shows the schematic geometry assuming a constant density for the ionized absorbers; the relative position and width of the ionized shells is roughly to scale according to Table \ref{windtable}, while the placement of the cold shell is determined by the radiative transfer, i.e., by whether the ionized shells see an unabsorbed (UN, top) or absorbed (AN, bottom) continuum. The right column shows the shell geometry if we adjust the density profile to make the ionized shells adjacent with a monotonic temperature/ionization profile. Note that hot shell is interior to the warm shell for both the AN and the UN models at constant density but exterior to the warm shell for the $x>0$ AN model. This inversion is a result of requiring a monotonically increasing temperature profile. See text for details.  \label{fig:geom}}

\end{figure*}
Instead, let us consider the possibility that the two ionized absorption zones are effectively contiguous; this requires them to have different densities. The various geometries are illustrated in Figure \ref{fig:geom}. Depending on whether the temperature and ionization increase inwards or outwards, we will infer a different density profile during the flare, but this is constrained by the measured column densities and ionization parameters. Suppose the density profile is given by \begin{equation}\label{eq:density}n(R)=n_{\rm i}\left(\frac{R}{R_{\rm i}}\right)^{-x},\end{equation} where $n_{\rm i}$ and $R_{\rm i}$ are the electron density and radius of the inner ionized absorber. If we assume the temperature profile varies monotonically, then the inner absorber must be cooler for the AN models and hotter for the UN models. 

Combining Equations \ref{eq:r} and \ref{eq:density} for a given time interval (where the luminosity is constant), we can write the relative position of the inner and outer absorption zones as \begin{equation}\label{eq:outer1}R_{\rm o}=R_{\rm i}\left(\frac{\xi_{\rm i}}{\xi_{\rm o}}\right)^{1/(2-x)}.\end{equation} To force the two zones to be contiguous, we specify \begin{equation}\label{eq:outer2}R_{\rm o}=R_{\rm i}+\Delta R_{\rm i},\end{equation}
where
\begin{eqnarray}\label{eq:nhdr1}N_{\rm H,i}&=&\int_{R_{\rm i}}^{R_{\rm o}}n(R)dR\\
&=&\int_{R_{\rm i}}^{R_{\rm 0}}n_{\rm i}\left(\frac{R}{R_{\rm i}}\right)^{-x}dR\label{eq:nhdr2}\\
&=&\frac{n_{\rm i}R_{\rm i}^x}{1-x}(R_{\rm o}^{1-x}-R_{\rm i}^{1-x}),\label{eq:nhdr3}
\end{eqnarray} assuming $x\neq1$. Solving for $R_{\rm o}$ for a given $N_{\rm H,i}$ gives 
\begin{equation}
    R_{\rm o}=\left(R_{\rm i}^{1-x}+\frac{N_{\rm H,i}(1-x)}{n_{\rm i}R_{\rm i}^x}\right)^{\frac{1}{1-x}}.\label{eq:outer3}
\end{equation} This formulation uses the {\sc xstar} convention that $R$ represents the inner edge of a shell of ionized gas and $\Delta R$ represents its width (as opposed to the center and half-width of the shell, respectively). Together, equations \ref{eq:outer2}- \ref{eq:outer3} allow us to use the measured column densities to calculate the radius of the outer shell, as long as we specify some initial density $n_{\rm i}$ and $x,$ the slope of the density profile.

By equating Equations \ref{eq:outer1} and \ref{eq:outer3}, we can find pairs of $n_{\rm i}$ and $x$ that make the inner and outer shells adjacent to one another. An exploration of these solutions reveals several important points:
\begin{enumerate}
     \item The density decreases outwards ($x>0$) for the absorbed models and increases outwards for the unabsorbed models ($x<0$).
   \item The inner density $n_{\rm i}$ cannot be much less than $10^{12}$ cm$^{-3},$ or the shell will be larger than the disk. This is due to our relatively low values of  $\xi$.
    \item As $n_{\rm i}$ increases, the shells become narrower ($R_{\rm o}$ approaches $R_{\rm i}$ in the high-density limit; see Equation \ref{eq:outer3}). Equation \ref{eq:outer1} has no density dependence, and so there no solutions for high densities. In practice, this means $n_{\rm i}$ cannot be much larger than $10^{13}$ cm$^{-3}$.
    \item Narrower shells require steeper density profiles (larger values of $|x|$) to satisfy Equation \ref{eq:outer1}. 
    \item If $x$ is too large (if $n(R)$ decreases too rapidly) or if the density is too low, it can be difficult to satisfy Equation \ref{eq:nhdr3}, especially for the outer shell. For this reason, the density profile cannot generally be a single power law over the interval $[R_{\rm i},R_{\rm disk}].$ 
\end{enumerate}

Remarkably, we arrive at a constraint on the density that is roughly independent of the slope of the density profile: $n_{\rm i}\sim10^{12}-10^{13}$ cm$^{-3}.$ The typical $R_{i}$ at these densities is a few$\times10^{11}$ cm for both models, although there is some variation from one interval to the next.  We do not believe there are large physical changes in the radius on short timescales. We interpret these changes (see also the changes in the derived radii in Table \ref{windtable}) as a reflection of variations in the density and structure of the absorbing medium, i.e., an imprint of its inhomogeneity or azimuthal asymmetry (Section \ref{sec:variability}). 

As far as the radial density profile, the UN models require $x$ between -3 and -1.7; the AN models have $x\sim2.6-11$. That is, in order to have an (AN) absorption geometry like the top right diagram in Figure \ref{fig:geom}, the density must drop sharply at the outer edge of the absorber; forcing the UN model to produce contiguous shells requires an outward increase in the density, albeit one that is less sharp than the decrease in the AN models.

In the dipping LMXB EXO 0748-676, \citet{Psaradaki18} found a two-zone absorber in eclipse-phase spectra from \textit{XMM-Newton}. They associated the cooler component with the clumpy outer disk and the warmer component with the disk atmosphere, closer to the neutron star. This matches the geometry of many two-zone absorbers in both dipping LMXBs and AGN (e.g., \citealt{Boirin04,Xiang09,Costantini16,Mehdipour17} and references therein) as well as the modeling by \citetalias{Miller20}. We therefore regard the UN model to be a priori more likely. On the other hand, the sharply increasing density profile at large radii required by the UN models seems less plausible, so we will not reject the absorbed (AN) models outright (they also have slightly lower $\chi^2$). In any case, many of the parameters derived from these models are similar. For instance, because the implied mass (Equation \ref{eq:m}) depends only on $R$ and $N_{\rm H}$, the inner shell masses are similar: $M_{\rm warm}\sim10^{22.5-23}$ g. The AN and UN models do, however, differ in the radial width of the shells: $\Delta R_{\rm i}/R_{\rm i}$ ranges from 0.05-0.6 for the UN models but from 0.2-10 for the AN models.

\subsection{Variability Timescales}
\label{sec:variability}
It is also possible to use the variability of the partial covering absorber to probe its geometry. Following \citet{Oosterbroek97}, we can consider a spherical cloud of cold material crossing the line of sight (or in our case, a cloud of material that has just finished crossing the line of sight). We suppose that the variability timescale is equal to the crossing time of the cloud: $t=l/v(R)$, where $l$ is the characteristic size of the cloud and $v(R)$ is its orbital speed, assumed to be Keplerian. For a uniform clump of density $n$, we can write the change in column density as $\Delta N_{\rm H}=n l\approx7\times10^{23}$ cm$^{-2}$. With Equation \ref{eq:xi} (and $\xi<\xi_{\rm max}\approx50$ for cold absorption; \citealt{Oosterbroek97} and references therein), we have a constraint on the radius or density of the cold absorber:
\begin{equation}
    R\gtrsim\left(\frac{L\sqrt{GM}t}{\xi_{\rm max} \Delta N_{\rm H}}\right)^{2/5}
\end{equation} and
\begin{equation}
    n\gtrsim\left(\frac{\Delta N_{\rm H}}{\sqrt{GM}t}\right)^{4/5}\left(\frac{L}{\xi_{\rm max}}\right)^{1/5}.
\end{equation} For the luminosities above and variability timescale $t=400$ s (roughly the time from the end of interval 1 to the end of interval 5), we find $R\gtrsim10^{11}$ cm and $n\gtrsim2\times10^{13}$ cm$^{-3}.$ 

The coincidence here is remarkable. Based on photoionization analysis, we find that the ionized absorption lines in GRS 1915+105 are most likely produced in hot gas of density $10^{12-13}$ cm$^{-3}$ at a distance of a few$\times10^{11}$ cm from the black hole.  Now our completely independent estimates based on the timescale for changes in the cold column density place the cold obscuration at similar radii and densities. 

For comparable timescales and radii in V404 Cyg, \citet{Oosterbroek97} argued that the variability could be attributed to the rim of the accretion disk. \citet{Morningstar14} considered a similar torus-like outer disk for V4641 Sgr. The size of the disk in GRS 1915+105 is similar to the orbital separation in V404 Cyg, so we conclude that the obscured state may represent a persistent, large-scale change in the structure of the outer accretion disk.

\subsection{Physical Origin of the Obscured State}
\label{sec:origin}
If the obscured state in GRS 1915+105 is caused by changes in the structure of the outer disk, what could be responsible for those changes? This remains a puzzle. As noted in Section \ref{sec:intro}, several other X-ray binaries (e.g., V404 Cyg, V4641 Sgr, and Swift J1538.6-014, in addition to numerous AGN) have shown evidence for significant changes in partial covering obscuration on relatively short timescales. In the X-ray binaries, these changes are often seen in the middle of bright outbursts and interpreted as signatures of extreme outflows launched at super-Eddington accretion rates (e.g., \citealt{Motta17b}), with variability driven by changing levels of obscuration. Meanwhile, several lines of evidence indicate that so-called ``changing-look" AGN are \textit{not} driven primarily by variable obscuration but instead by intrinsic changes in the accretion power or state \citep{Noda18,Dexter19}.

There are several possible explanations. A super-Eddington outflow like the one described by \citet{Motta17b} in V404 Cyg could produce obscuration (and potentially the continuum emission itself, but this requires caution when interpreting radii inferred from photoionization models; see \citealt{N16}). But the origin of the outflow in this scenario is still unclear. Why, after spending decades accreting at near-Eddington rates and producing massive winds (e.g., \citealt{N11a}), did GRS 1915+105 suddenly start to produce a wind capable of obscuring nearly the entire continuum emission? State changes are certainly not unusual in GRS 1915 \citep{b00}, but this explanation leaves several questions unanswered.

We can also consider a scenario where the secondary star is directly or indirectly responsible for the obscuration, whether due to irradiation, wind, the impact point between the disk and the accretion stream, or a change in the mass supply rate (see below). Irradiation does not seem likely because the apparent luminosity of the accretion flow is quite low in the obscured state, so the companion should be less irradiated than usual. Similarly, winds and changes in the accretion stream hot spot do not seem particularly likely. It is possible that the secondary has activity cycles on timescales of decades that might affect its stellar wind or the accretion stream, but these do not seem likely to produce the required changes in the disk. If the obscuration is related to the accretion stream or secondary star, one might expect to detect an orbital phase dependence in the properties of the obscuration (see also discussion in \citealt{Morningstar14}). We will explore this possibility in future work.

Finally, there is the possibility of structural changes in the outer accretion disk. As noted in Section \ref{sec:variability}, the outer rim of the disk itself may be responsible for the obscuration. If this is the case, the timescale over which the scale height of the disk can change is the thermal timescale $t_{\rm th}\simeq (\alpha\Omega)^{-1}$, where $\alpha$ is the viscosity parameter and $\Omega$ is the Keplerian orbital angular velocity. Assuming $\alpha=0.01$, for a 12.6 $M_{\odot}$ black hole we find $t_{\rm th}\sim$5-150 days for distances from $3-30\times10^{11}$ cm. In other words, the thermal timescale in the outer disk is comparable to the time it took for GRS 1915+105 to enter the obscured state. 

But there is still the lingering question of why any such changes might have happened in the first place. Irradiation can increase the scale height of the outer disk, which raises the possibility that changes in the inner disk have led to enhanced irradiation, resulting in changes in the outer disk (a common causal relationship with disk winds; \citealt{N11a,N12a}. But again, the decrease in the apparent flux makes irradiation-related explanations puzzling. Another possibility is the impending end of this 28-year outburst. In one version of the disk instability model as applied to X-ray transients (see \citealt{Lasota01,Dubus01} and references therein), outbursts begin in the outer disk and propagate inward until the reservoir of excess matter in the outer disk is depleted: both the beginning and the end proceed outside-in\footnote{This represents a difference from changing-look quasars, where \citet{Dexter19} suggested that disk instabilities might be responsible for transitions as long as they occur inside-out.}. In effect, the transition toward quiescence may begin with structural changes in the outer disk. On the other hand, cooling the outer disk should decrease its scale height, so the viability of approaching quiescence as an explanation for the obscured state depends on the exact details of the irradiation and temperature profile of the disk (see \citealt{Dubus01}). In any case, the viscous timescale is so long in the extended disk of GRS 1915+105 that it may be possible for us to watch changes in the disk structure play out over a relatively long time (compared to more compact systems).

It would be exciting if the onset of the obscured state was the first X-ray signature of quiescence in GRS 1915+105, but regardless it appears to indicate a change in the scale height of the outer disk that resulted in irregular obscuration of the central X-ray source. The black hole will continue accreting at its usual clip until any changes in the outer disk accretion rate propagate to the inner edge of the disk. If those changes lead to quiescence, we will see the source disappear. But it is worth noting that the latter half of the obscured state in Figure \ref{fig:lc} appears to exhibit more correlated variability than the first half; this could just as easily be a signature of the end of the obscured state. Indeed, an \textit{increase} in the mass supply rate from the companion could result in strong outflows or changes in the outer disk that might lead to obscuration. While the appropriate slim disk solution at high accretion rate would be less radiatively efficient than a thin disk (e.g., \citealt{Mineshige00} and references therein), it is not clear if this is sufficient to explain why there has been no long-term change in the intrinsic/unabsorbed luminosity of the source. It is also unclear whether the mass supply rate from the companion star could reasonably increase enough on short enough timescales to induce such a state. In either case, it seems wise to continue monitoring this important system to discern the relationship between the obscuring medium and the intrinsic luminosity of the inner accretion flow.

\subsubsection{Comparison to \citetalias{Miller20}}

\citetalias{Miller20} observed GRS 1915+105 three times in 2019 with the \textit{Chandra} gratings. Their two latter observations deep in the obscured state (MJD 58790.37, 58817.36) revealed spectra similar to our interval 1, with a flat continuum and narrow emission lines. They found that the line regions were were well described by a combination of cold reflection and hot emission, all likely far from the black hole. In their first observation, taken during the onset of obscured state, they detected strong absorption lines. It is enlightening to compare their results to ours here despite the different behavior of the underlying accretion flow. Since our primary interest in this work is the absorption during our flare, we focus here on their analysis of their absorption-line spectrum. 

As echoed by our \nicer data, \citetalias{Miller20} required a two-zone absorber, with a faster outflowing inner component and a $\sim$stationary outer region. The \textit{Chandra} gratings have superior energy resolution, but this is qualitatively very similar to what we have found during the flare. Our hot absorption component is generally more blueshifted than the cooler component, and we have interpreted it as being likely closer to the black hole. The primary difference in our direct results is that their ionized column densities are higher than ours and their ionization parameters are higher (a reflection of the different shapes of the ionizing spectra). For a fixed density, this would imply material closer to the black hole. We also make different assumptions about the absorption geometry: \citetalias{Miller20} make the common assumption that $\Delta R\approx R$, which leads to a lower density and larger radius than would be inferred from our Equations \ref{eq:r} and \ref{eq:dr}. 

Finally, \citetalias{Miller20} regard their two absorption components as possibly having different origins (due in part to the superior ability of the gratings to detect and infer significant blueshifts and redshifts). The inner fast component is, in their model, a failed (bound) magnetic wind; the outer cool component might be a thermally-driven outflow. They suggest that the failed wind is the source of the obscuration in this state. A magnetic mechanism might help to explain the onset of the obscured state without a large change in the intrinsic luminosity of the accretion flow. But the velocity of the failed wind that they infer is not large enough for the gas to reach radii comparable to the location of the cold gas in our analysis, so the relationship between these components is unclear. With \nicer's gain uncertainty and velocity resolution, we cannot rule out a failed wind scenario, but our lower ionization parameters make it more straightforward to place both zones in the outer disk where the gas could escape to infinity, and where a single mechanism (e.g., a radially-stratified, puffed up outer disk; Figure \ref{fig:geom}) could be responsible for the hot, warm, and cold obscuration present in our data. 

\subsection{Model Dependence}
\label{sec:assumptions}
Finally it is important to discuss the limitations of our model. We relied on the similarity of our spectra to spectra of V404 Cyg and other sources (e.g., \citealt{Motta17b}) to motivate a partial covering model. The implication is that the flat underlying portion of the spectra is uncovered, while the bright $>3$ keV excess (Section \ref{sec:spec}) is obscured by a high column density absorber. 

One curiosity of this model is the strong correlation between the continuum luminosity and the partial covering fraction (correlation coefficient $r>0.9$). We have a small number of data points, but in principle one would expect these quantities to be independent, especially if the obscurer is far from the black hole (Section \ref{sec:variability}). As an alternative, we could treat the continuum as the sum of two distinct parts: a highly obscured component (the $>3$ keV excess) and a scattered component (the flat underlying continuum and narrow emission lines). This model is similar to the one used by \citetalias{Miller20}. In this scenario, the scattered emission is fairly steady but briefly rises (interval 2), while the obscured component is responsible for the flare itself. However, because the underlying unobscured continuum does not rise significantly during the brightest portions of the flare, this scenario also requires some fine-tuning (i.e., changes in the visibility of the scattering region). That is, why does the increase in the luminosity during the flare not result in a significant increase in the scattered component? These two scenarios thus require similar amounts of fine tuning without significant differences in their model formulations or implications, so we regard them as equivalent.

Another component of the model worth discussing is the broad iron line. We have argued above that the emitting region must be compact, but given uncertainty in the nature of the obscured state, it is not certain that we have a direct line of sight to the inner accretion disk. The main utility of the broad line in the fits appears to be capturing the shallow slope of the 7 keV edge, especially in interval 3 (where the inferred inner radius is the lowest). Eliminating the broad line but allowing the widths and energies of the three narrow emission lines to vary results in a fit to the interval 3 spectrum that is almost as good ($\Delta \chi^2\approx6$ for 3 fewer parameters). So it is possible that the shallower edge could be the blue wing of a non-relativistic emission line. This would more closely track the results of \citetalias{Miller20}, who found that the continuum leading into the obscured state was well described with an absorbed, moderately-blurred reflection component. In addition, \citet{Motta17a} demonstrate that the inferences drawn from spectral models in the Compton thick regime are sensitive to the precise assumptions made. Interval 3 in particular has the highest total column density, and it is possible that the simple model we have used here does not capture the full complexity of radiative transfer in a high optical depth obscuring medium. This might explain why many of our line energies fall below $\sim$6 keV. However, given the broad similarity of our results to those of \citetalias{Miller20}, we do not believe a more complex model would substantially change our interpretation.

\section{Conclusion}
\label{sec:concl}

In 2018 and 2019, the apparent X-ray brightness of GRS 1915+105 dropped precipitously after a long and luminous outburst. The source entered a faint but variable state that is punctuated occasionally by bright flares. Our spectral modeling of one such flare observed by \nicer indicates that despite appearances, the intrinsic luminosity of GRS 1915+105 is still large: the 2--10 keV luminosities of $(5-20)\times10^{37}$ erg s$^{-1}$ before and during the flare are comparable to those measured by \nicer when the source was in a much more visibly active state \citep{n18}. 

Instead, there are clear indications that the drop in flux is related to obscuration of the X-ray emission. Our \nicer spectra (Figure \ref{fig:spec}) indicate the complexity of this obscuration. Based on our analysis with {\sc xstar}, the obscuring medium is inhomogeneous and requires at least three different temperature zones (see Figure \ref{fig:geom}): one high column density cold zone to describe the soft X-ray absorption and two highly ionized zones to model the rich, strong absorption line spectrum. The ionized absorption appears to be in motion, but given the size of our statistical and systematic uncertainties, the speed and mass loss rate of a sustained outflow are unclear.

Assuming the X-ray source is compact (which is justified by the small size of the emitting region inferred from blackbody continuum models), we use ionization and variability arguments to probe the geometry and properties of the obscuring cloud. By requiring that the two ionized zones be physically adjacent, we find that the partial covering medium can be described as radially stratified, with a density of $10^{12-13}$ cm$^{-3}$ at a radius of a few$\times10^{11}$ cm from the black hole. 

While high optical depth partial covering absorption is not uncommon in the outbursts of stellar-mass black holes and is often attributed to massive outflows (\citealt{Zycki99,Oosterbroek97,Revnivtsev02,Morningstar14,Motta17a,Motta17b,Walton17,Hare20,Munoz-Darias20}), the large radii in question here are also consistent with the outer accretion disk. It is possible that the scale height of the outer disk has increased sufficiently to partially obscure the inner accretion flow. We consider several scenarios to explain this, from an increase in mass accretion rate to global changes to the disk structure leading toward quiescence, but none are entirely satisfactory. Continuing to monitor GRS 1915+105 and correlating the intrinsic spectral evolution of the accretion flow with the properties of this new obscuring medium may reveal more about its origin and its trajectory.

\acknowledgements
We thank Zaven Arzoumanian for his assistance scheduling and performing the observations. This research was supported by NASA Award 80NSSC19K1456. This research has made use of data, software and/or web tools obtained from the High Energy Astrophysics Science Archive Research Center (HEASARC), a service of the Astrophysics Science Division at NASA/GSFC and of the Smithsonian Astrophysical Observatory's High Energy Astrophysics Division. This research has made use of a collection of ISIS functions (ISISscripts) provided by ECAP/Remeis observatory and MIT (http://www.sternwarte.uni-erlangen.de/isis/).
\bibliographystyle{aasjournal}
\bibliography{ms}

\end{document}